\documentclass[english,jcp,preprint,endfloats,groundaddress]{revtex4-1} 
\usepackage[T1]{fontenc}
\usepackage{graphicx}
\usepackage{amsfonts}
\usepackage{amssymb}
\usepackage{longtable}
\usepackage{multirow}
\usepackage{supertabular}
\usepackage{epsfig}

\makeatletter

\begin{document}

\title{Thermally-assisted-occupation density functional theory with generalized-gradient approximations} 
\author{Jeng-Da Chai} 
\email{Electronic mail: jdchai@phys.ntu.edu.tw} 
\address{Department of Physics, Center for Theoretical Sciences, and Center for Quantum Science and Engineering, National Taiwan University, Taipei 10617, Taiwan} 

\date{\today{}}

\begin{abstract} 

We extend the recently proposed thermally-assisted-occupation density functional theory (TAO-DFT) [J.-D. Chai, J. Chem. Phys. {\bf 136}, 154104 (2012)] to generalized-gradient approximation (GGA) 
exchange-correlation density functionals. Relative to our previous TAO-LDA (i.e., the local density approximation to TAO-DFT), the resulting TAO-GGAs are significantly superior for a wide range of applications, 
such as thermochemistry, kinetics, and reaction energies. For noncovalent interactions, TAO-GGAs with empirical dispersion corrections are shown to yield excellent performance. 
Due to their computational efficiency for systems with strong static correlation effects, TAO-LDA and TAO-GGAs are applied to study the electronic properties (e.g., the singlet-triplet energy gaps, 
vertical ionization potentials, vertical electron affinities, fundamental gaps, and symmetrized von Neumann entropy) of acenes with different number of linearly fused benzene rings (up to 100), which is very challenging 
for conventional electronic structure methods. The ground states of acenes are shown to be singlets for all the chain lengths studied here. With the increase of acene length, the singlet-triplet energy gaps, 
vertical ionization potentials, and fundamental gaps decrease monotonically, while the vertical electron affinities and symmetrized von Neumann entropy (i.e., a measure of polyradical character) increase monotonically. 

\end{abstract}

\maketitle
\newpage

\section{Introduction} 

Due to its favorable balance between cost and performance, Kohn-Sham density functional theory (KS-DFT) \cite{HK,KS} has been a very popular method for the study of large ground-state 
systems \cite{Parr,DFTReview,Perdew09}. However, its essential ingredient, the exact exchange-correlation (XC) density functional $E_{xc}[\rho]$, remains unknown and needs to be approximated. 
Functionals based on the conventional density functional approximations (DFAs), such as the local density approximation (LDA) and generalized gradient approximations (GGAs), are reliably accurate 
for properties governed by short-range XC effects, and are computationally efficient for very large systems. However, KS-DFAs can produce erroneous results in situations where the nonlocal XC effects 
are pronounced. Over the past two decades, numerous attempts have been made to resolve the qualitative failures of KS-DFAs \cite{DFTReview,Perdew09,SciYang,RevYang,TAO-DFT}. 

Recently, we have shown that long-range corrected (LC) hybrid functionals \cite{LC-DFT,LCHirao,CAM-B3LYP,LC-wPBE,BNL,wB97X,wB97X-D,Herbert,wB97X-2,M11,wM05-D,LC-D3}, incorporating the 
long-range Hartree-Fock (HF) exchange into the KS-DFAs, could be reliably accurate for a very wide range of applications \cite{LCAC}, especially for those sensitive to the long-range HF exchange, 
such as the asymptote problems \cite{exactip,asymp,LB94,sum_vxc2,sum_vxc3,Staroverov,AC1,LFA}, self-interaction-error problems \cite{SIC,SIE}, 
energy-gap problems \cite{DD1,DD2,DD2a,DD4a,DDcorr,DD6,DD7,errsinfuncs,DD8,DDHirao,GKS2,Correction2,Correction3,DDChai}, and charge-transfer problems \cite{Dreuw,Tozer,Gritsenko,Dreuw2,Dreuw3,fxcd}. 

To reduce the qualitative failures of KS-DFAs for noncovalent interactions \cite{Dobson2}, the DFT-D (KS-DFT with empirical dispersion corrections) schemes \cite{DFT-D1,DFT-D2,wB97X-D,DFT-D3}, adding 
empirical atom-atom dispersion potentials into the KS-DFAs, have shown generally satisfactory performance on a large set of noncovalent systems \cite{Sherrill,DFT-D3appl}. Alternatively, the 
double-hybrid (DH) methods \cite{B2PLYP,XYG3,wB97X-2,ST2011,PBE0-DH,TS2011,PBE0-2,lrc-XYG3,OO-PBE0-2}, mixing some of the HF exchange and some of the nonlocal orbital correlation energy 
from the second-order M\o ller-Plesset perturbation (MP2) theory \cite{MP2} into the KS-DFAs, can also be adopted to take into account nonlocal dynamical correlation effects. DH functionals have shown 
an overall satisfactory accuracy for thermochemistry, kinetics, noncovalent interactions, and self-interaction-error problems. 

Despite their computational efficiency, KS-DFAs, global hybrid functionals \cite{hybrid}, LC hybrid functionals, and DH functionals perform very poorly for strongly correlated (SC) systems (i.e., multi-reference systems), 
which are systems with strong static correlation effects, including bond-breaking reactions, conjugated polymers, and transition-metal compounds \cite{SciYang,RevYang,TAO-DFT,SCE}. Within the framework of KS-DFT, 
fully nonlocal XC functionals, such as those based on random phase approximation (RPA), can be essential for the accurate treatment of SC systems. However, RPA-type functionals are computationally very demanding 
for large systems \cite{DFTReview,Perdew09,RPA_Yang,RPA_Yang2}. 

Aiming to reduce the static-correlation-error problems with minimum computational complexity, we have recently developed thermally-assisted-occupation DFT (TAO-DFT) \cite{TAO-DFT}, a DFT with fractional orbital 
occupations given by the Fermi-Dirac distribution (controlled by a fictitious temperature $\theta$), for the study of large systems with strong static correlation effects. In contrast to finite-temperature DFT \cite{Mermin}, 
TAO-DFT is developed for ground-state systems at zero temperature. TAO-DFT has similar computational cost as KS-DFT, and is reduced to KS-DFT in the absence of strong static correlation effects. Even at the simplest 
LDA level, the resulting TAO-LDA has been shown to consistently improve upon KS-LDA for multi-reference systems. However, TAO-LDA performs similarly to KS-LDA for single-reference systems, due to the absence of 
strong static correlation. 

To improve the performance of TAO-LDA for single-reference systems, here we propose TAO-GGAs for the improved description of short-range XC effects. 
Relative to TAO-LDA, the resulting TAO-GGAs are significantly superior for a wide range of applications, such as thermochemistry, kinetics, and reaction energies. 
For noncovalent interactions, TAO-GGAs with empirical dispersion corrections are shown to yield excellent performance. The rest of this paper is organized as follows. 
In section II, we briefly describe the formulation of TAO-DFT and the DFAs to TAO-DFT. The performance of TAO-LDA and TAO-GGAs is compared with that of KS-LDA and KS-GGAs in section III. 
In section IV, we apply TAO-LDA and TAO-GGAs to study the electronic properties of linear acenes. Our conclusions are given in section V.

\section{TAO-DFT} 

For a system with $N_{\alpha}$ up-spin electrons and $N_{\beta}$ down-spin electrons in an external potential $v_{ext}({\bf r})$ at zero temperature, two noninteracting auxiliary systems at the same fictitious temperature 
$\theta$ (measured in energy units) are adopted in spin-polarized (spin-unrestricted) TAO-DFT \cite{TAO-DFT}: one described by the spin function $\alpha$ and the other by function $\beta$, with the respective thermal equilibrium 
density distributions $\rho_{s,\alpha}({\bf r})$ and $\rho_{s,\beta}({\bf r})$ exactly equal to $\rho_{\alpha}({\bf r})$ and $\rho_{\beta}({\bf r})$, respectively, in the original interacting system at zero temperature. 
The resulting self-consistent equations for $\sigma$-spin electrons ($\sigma$ = $\alpha$ or $\beta$) are given by ($i$ runs for the orbital index): 
\begin{equation} 
\bigg\lbrace -\frac{\hbar^2}{2 m_e} {\bf \nabla}^2 \ + \ v_{s,\sigma}({\bf r}) \bigg\rbrace \psi_{i,\sigma}({\bf r}) = \epsilon_{i,\sigma} \psi_{i,\sigma}({\bf r}), 
\label{eq:taodft1}
\end{equation} 
where the effective potential is 
\begin{equation} 
v_{s,\sigma}({\bf r}) = v_{ext}({\bf r}) + e^2 \int \frac{\rho({\bf r'})}{|{\bf r} - {\bf r'}|}d{\bf r'} + \frac{\delta E_{xc}[\rho_{\alpha},\rho_{\beta}]}{\delta \rho_{\sigma}({\bf r})} 
+ \frac{\delta E_{\theta}[\rho_{\alpha},\rho_{\beta}]}{\delta \rho_{\sigma}({\bf r})}. 
\label{eq:taodft2}
\end{equation} 
Here $E_{xc}[\rho_{\alpha},\rho_{\beta}]$ is the XC energy defined in spin-polarized KS-DFT \cite{SDFT,SDFTYang}, and 
$E_{\theta}[\rho_{\alpha},\rho_{\beta}] \equiv A_{s}^{\theta=0}[\rho_{\alpha},\rho_{\beta}] - A_{s}^{\theta}[\rho_{\alpha},\rho_{\beta}]$ 
is the difference between the noninteracting kinetic free energy at zero temperature and that at the fictitious temperature $\theta$. The $\sigma$-spin density is given by 
\begin{equation}
\rho_{\sigma}({\bf r}) = \sum_{i=1}^{\infty} f_{i,\sigma} |\psi_{i,\sigma}({\bf r}) |^{2},
\label{eq:taodft3}
\end{equation}
where the occupation number $f_{i,\sigma}$ is the Fermi-Dirac function
\begin{equation}
f_{i,\sigma} = \{1+\text{exp}[( \epsilon_{i,\sigma} - \mu_{\sigma})/ \theta] \}^{-1},
\label{eq:taodft4}
\end{equation}
and the chemical potential $\mu_{\sigma}$ is chosen to conserve the number of $\sigma$-spin electrons $N_{\sigma}$, 
\begin{equation}
\sum_{i=1}^{\infty} \{1+\text{exp}[( \epsilon_{i,\sigma} - \mu_{\sigma})/ \theta] \}^{-1} = N_{\sigma}. 
\label{eq:taodft5}
\end{equation}
The ground-state density $\rho({\bf r})$ is computed as the sum of the up-spin density $\rho_{\alpha}({\bf r})$ and down-spin density $\rho_{\beta}({\bf r})$: 
\begin{equation}
\rho({\bf r}) = \rho_{\alpha}({\bf r}) + \rho_{\beta}({\bf r}) = \sum_{\sigma=\alpha,\beta} \rho_{\sigma}({\bf r}). 
\label{eq:taodft6}
\end{equation} 
The formulation of spin-polarized TAO-DFT has yielded two sets (one for each spin function) of self-consistent equations, Eqs.\ (\ref{eq:taodft1}), (\ref{eq:taodft2}), (\ref{eq:taodft3}), (\ref{eq:taodft4}), and (\ref{eq:taodft5}), 
for $\rho_{\alpha}({\bf r})$ and $\rho_{\beta}({\bf r})$, respectively, which are coupled with $\rho({\bf r})$ by Eq.\ (\ref{eq:taodft6}). 

The self-consistent procedure described in Ref.\ \cite{TAO-DFT} may be adopted to obtain the spin densities and ground-state density. After self-consistency is achieved, the noninteracting kinetic free energy 
$A_{s}^{\theta}$ is given by 
\begin{equation} 
A_{s}^{\theta}[\{f_{i,\alpha}, \psi_{i,\alpha} \}, \{f_{i,\beta}, \psi_{i,\beta} \}] = T_{s}^{\theta}[\{f_{i,\alpha}, \psi_{i,\alpha} \}, \{f_{i,\beta}, \psi_{i,\beta} \}] + E_{S}^{\theta}[\{f_{i,\alpha} \}, \{f_{i,\beta} \}], 
\label{eq:taodft7}
\end{equation} 
which is the sum of the kinetic energy 
\begin{eqnarray} 
T_{s}^{\theta}[\{f_{i,\alpha}, \psi_{i,\alpha} \}, \{f_{i,\beta}, \psi_{i,\beta} \}] 
&=& -\frac{\hbar^2}{2 m_e} \sum_{\sigma=\alpha,\beta} \sum_{i=1}^{\infty} f_{i,\sigma} \int \psi_{i,\sigma}^{*}({\bf r}){\bf \nabla}^2\psi_{i,\sigma}({\bf r}) d{\bf r} \nonumber \\ 
&=& \sum_{\sigma=\alpha,\beta} \bigg\lbrace \sum_{i=1}^{\infty} f_{i,\sigma} \epsilon_{i,\sigma} - \int \rho_{\sigma}({\bf r}) v_{s,\sigma}({\bf r}) d{\bf r} \bigg\rbrace 
\label{eq:taodft8}
\end{eqnarray}
and entropy contribution 
\begin{equation} 
E_{S}^{\theta}[\{f_{i,\alpha} \}, \{f_{i,\beta} \}] = \theta \sum_{\sigma=\alpha,\beta} \sum_{i=1}^{\infty} \bigg\lbrace f_{i,\sigma}\ \text{ln}(f_{i,\sigma}) + (1-f_{i,\sigma})\ \text{ln}(1-f_{i,\sigma}) \bigg\rbrace 
\label{eq:taodft9}
\end{equation} 
of noninteracting electrons at the fictitious temperature $\theta$. The total ground-state energy $E[\rho_{\alpha},\rho_{\beta}]$ in spin-polarized TAO-DFT is evaluated by 
\begin{equation}
E[\rho_{\alpha},\rho_{\beta}] = A_{s}^{\theta}[\{f_{i,\alpha}, \psi_{i,\alpha} \}, \{f_{i,\beta}, \psi_{i,\beta} \}] + \int \rho({\bf r}) v_{ext}({\bf r}) d{\bf r} + E_{H}[\rho] 
+ E_{xc}[\rho_{\alpha},\rho_{\beta}] + E_{\theta}[\rho_{\alpha},\rho_{\beta}], 
\label{eq:taodft10}
\end{equation} 
where $E_{H}[\rho] \equiv \frac{e^2}{2} \int\int \frac{\rho({\bf r})\rho({\bf r'})}{|{\bf r} - {\bf r'}|}d{\bf r}d{\bf r'}$ is the Hartree energy. 
Spin-unpolarized (spin-restricted) TAO-DFT can be formulated by imposing the constraints of $\psi_{i,\alpha}({\bf r})$ = $\psi_{i,\beta}({\bf r})$ and $f_{i,\alpha}$ = $f_{i,\beta}$ to spin-polarized TAO-DFT. 

In spin-polarized TAO-DFT, as the exact $E_{xc}[\rho_{\alpha},\rho_{\beta}]$ and $E_{\theta}[\rho_{\alpha},\rho_{\beta}]$, in terms of the spin densities $\rho_{\alpha}({\bf r})$ and $\rho_{\beta}({\bf r})$, have not been known, 
DFAs for both of them (denoted as TAO-DFAs) are needed for practical applications. Accordingly, the performance of TAO-DFAs depends on the accuracy of DFAs and the choice of the fictitious temperature $\theta$. While 
$E_{xc}^{\text {DFA}}[\rho_{\alpha},\rho_{\beta}]$ can be readily obtained from those in spin-polarized KS-DFT, $E_{\theta}^{\text {DFA}}[\rho_{\alpha},\rho_{\beta}]$ can be obtained with the knowledge of 
$A_{s}^{\text{DFA}, \theta}[\rho_{\alpha},\rho_{\beta}]$ as follows: 
\begin{equation}
E_{\theta}^{\text {DFA}}[\rho_{\alpha},\rho_{\beta}] \equiv A_{s}^{\text{DFA}, \theta=0}[\rho_{\alpha},\rho_{\beta}] - A_{s}^{\text{DFA}, \theta}[\rho_{\alpha},\rho_{\beta}]. 
\label{eq:taodft11}
\end{equation}
Note that $E_{\theta=0}^{\text {DFA}}[\rho_{\alpha},\rho_{\beta}] = 0$ (i.e., an exact property of $E_{\theta}[\rho_{\alpha},\rho_{\beta}]$) is ensured by Eq.\ (\ref{eq:taodft11}). From the spin-scaling relation of 
$A_{s}^{\theta}[\rho_{\alpha},\rho_{\beta}]$ \cite{spin-scaling}, Eq.\ (\ref{eq:taodft11}) can be expressed in terms of $A_{s}^{\text{DFA}, \theta}[\rho]$ (in its spin-unpolarized form): 
\begin{equation}
E_{\theta}^{\text {DFA}}[\rho_{\alpha},\rho_{\beta}] 
= \frac{1}{2} (A_{s}^{\text{DFA}, \theta=0}[2\rho_{\alpha}] + A_{s}^{\text{DFA}, \theta=0}[2\rho_{\beta}]) - \frac{1}{2} (A_{s}^{\text{DFA}, \theta}[2\rho_{\alpha}] + A_{s}^{\text{DFA}, \theta}[2\rho_{\beta}]). 
\label{eq:taodft12} 
\end{equation}

In our previous work \cite{TAO-DFT}, Perrot's parametrization of $A_{s}^{\text{LDA}, \theta}[\rho]$, which is the LDA for $A_{s}^{\theta}[\rho]$ (see Appendix A of Ref.\ \cite{As}), was adopted to obtain 
$E_{\theta}^{\text {LDA}}[\rho_{\alpha},\rho_{\beta}]$. To go beyond the simple LDA, $A_{s}^{\text{GEA}, \theta}[\rho]$, which is the gradient expansion approximation (GEA) for $A_{s}^{\theta}[\rho]$ 
(see Appendices A and B of Ref.\ \cite{As}), can be adopted to obtain $E_{\theta}^{\text {GEA}}[\rho_{\alpha},\rho_{\beta}]$. For the nearly uniform electron gas, $E_{\theta}^{\text {GEA}}[\rho_{\alpha},\rho_{\beta}]$ is 
expected to improve upon $E_{\theta}^{\text {LDA}}[\rho_{\alpha},\rho_{\beta}]$. 

As discussed in Ref.\ \cite{TAO-DFT}, TAO-DFT offers an explicit description of strong static correlation via the entropy contribution $E_{S}^{\theta}[\{f_{i,\alpha} \}, \{f_{i,\beta} \}]$ (see Eq.\ (\ref{eq:taodft9})). Even at the 
simplest LDA level, the resulting TAO-LDA has been shown to perform reasonably well for multi-reference systems (due to the appropriate treatment of static correlation), when the distribution of orbital occupation 
numbers $\{f_{i,\sigma}\}$ (related to the chosen $\theta$) is close to that of the natural orbital occupation numbers (NOONs) \cite{NO}. This implies that a system-dependent $\theta$ (related to the distributions of NOONs) 
should be needed to capture the essential physics of strong static correlation effects. However, for the sake of simplicity, an optimal value of $\theta$ = 7 mhartree was previously defined for TAO-LDA, based on physical arguments 
and numerical investigations. Interestingly, TAO-LDA (with $\theta$ = 7 mhartree) was shown to consistently improve upon KS-LDA for multi-reference systems, while performing similarly to KS-LDA for single-reference systems. 
As TAO-GGAs should improve upon TAO-LDA mainly for properties governed by short-range XC effects, the optimal values of $\theta$ for TAO-GGAs should be similar to that for TAO-LDA, when the same physical arguments 
and numerical investigations are employed to define the optimal $\theta$. Therefore, we adopt an optimal value of $\theta$ = 7 mhartree for all the TAO-LDA and TAO-GGAs calculations in this work, unless noted otherwise. 
The limiting case where $\theta = 0$ for TAO-DFA is especially interesting, as this reduces to KS-DFA. Therefore, it is important to know how well KS-DFA performs here to assess the significance of the extra parameter 
$\theta$ for TAO-DFA.

\section{Results and Discussion for the Test Sets} 

For a comprehensive comparison, we examine the performance of various XC functionals: 
LDA \cite{LDA} and three popular GGAs (PBE \cite{PBE}, BLYP \cite{BLYP}, and BLYP-D \cite{BLYP-D}) in both KS-DFT and TAO-DFT, on various test sets involving 
the 223 atomization energies (AEs) of the G3/99 set \cite{G2a}, the 40 ionization potentials (IPs), 25 electron affinities (EAs), and 8 proton affinities (PAs) of the G2-1 set \cite{G21}, 
the 76 barrier heights (BHs) of the NHTBH38/04 and HTBH38/04 sets \cite{BH}, the 22 noncovalent interactions of the S22 set \cite{S22}, 
the reaction energies of the 30 chemical reactions in the NHTBH38/04 and HTBH38/04 sets \cite{BH}, the 166 optimized geometry properties of EXTS \cite{EXTS}, and two dissociation curves for H$_2$ and N$_2$. 
There are in total 592 pieces of data in our test sets, which are very large and diverse. Detailed information about the test sets may be found in Ref.\ \cite{wB97X}. 

For all the TAO-DFT calculations, $\theta$ = 7 mhartree is adopted, unless noted otherwise. Our preliminary TAO-DFT results show that the difference between using $E_{\theta}^{\text {LDA}}$ and $E_{\theta}^{\text {GEA}}$ 
is much smaller than the difference between using two different XC functionals. Unsurprisingly, as $E_{\theta=0}^{\text {LDA}} = E_{\theta=0}^{\text {GEA}} = 0$, the difference between $E_{\theta}^{\text {LDA}}$ 
and $E_{\theta}^{\text {GEA}}$ should remain small for a sufficiently small $\theta$ (i.e., 7 mhartree). Therefore, for brevity, we only adopt $E_{\theta}^{\text {LDA}}$ for the TAO-DFT calculations in this work. 

All calculations are performed with a development version of \textsf{Q-Chem 4.0} \cite{QChem}. Spin-restricted theory is used for singlet states and spin-unrestricted theory for others, unless noted otherwise. 
Results for the test sets are computed using the 6-311++G(3df,3pd) basis set with the fine grid EML(75,302), consisting of 75 Euler-Maclaurin radial grid points \cite{EM} and 302 Lebedev angular grid points \cite{L}. 
For the interaction energies of the weakly bound systems, the counterpoise correction \cite{CP} is employed to reduce the basis set superposition error (BSSE). 
The error for each entry is defined as (error = theoretical value $-$ reference value). The notation used for characterizing statistical errors is as follows: 
mean signed errors (MSEs), mean absolute errors (MAEs), root-mean-square (rms) errors, maximum negative errors (Max($-$)), and maximum positive errors (Max(+)).

\subsection{$\omega$B97 training set} 

The $\omega$B97 training set \cite{wB97X} contains several well-known databases, such as the 223 AEs of the G3/99 set \cite{G2a}, the 40 IPs, 25 EAs, and 8 PAs of the G2-1 set \cite{G21}, 
the 76 BHs of the NHTBH38/04 and HTBH38/04 sets \cite{BH}, and the 22 noncovalent interactions of the S22 set \cite{S22}. Table \ref{table:training} summarizes the statistical errors of 
various functionals in both KS-DFT and TAO-DFT for the $\omega$B97 training set. As shown, TAO-DFAs perform comparably to the corresponding KS-DFAs. 
Unsurprisingly, as these systems do not have much static correlation, the exact NOONs should be close to either 0 or 1, which can be well simulated by the orbital occupation numbers of TAO-DFAs 
(with a sufficiently small $\theta$ = 7 mhartree). Due to the improved treatment of short-range XC effects, TAO-GGAs significantly outperform TAO-LDA for the AEs of the G3/99 set, 
the PAs of the G2-1 set, and the BHs of the NHTBH38/04 and HTBH38/04 sets \cite{supp}. For the IPs and EAs of the G2-1 set, TAO-GGAs perform slightly better than TAO-LDA. 
For the noncovalent interactions of the S22 set, the dispersion corrected functionals (i.e., KS-BLYP-D and TAO-BLYP-D) are found to be very accurate, while all the other functionals perform poorly \cite{supp}. 
This suggests that for noncovalent interactions, the successful DFT-D schemes for KS-DFT remain very accurate for TAO-DFT.

\subsection{Reaction energies} 

The reaction energies of the 30 chemical reactions (those with different barrier heights for the forward and backward directions) taken from the NHTBH38/04 and HTBH38/04 sets, are used to examine the performance of 
KS-DFAs and TAO-DFAs. As shown in Table \ref{table:reall}, TAO-DFAs have very similar performance to the corresponding KS-DFAs. TAO-GGAs are shown to significantly improve upon TAO-LDA for this test set \cite{supp}.

\subsection{Equilibrium geometries}

Satisfactory predictions of molecular geometries are essential for practical applications. Geometry optimizations for various functionals in both KS-DFT and TAO-DFT are performed on the 
equilibrium experimental test set (EXTS) \cite{EXTS}, consisting of 166 symmetry unique experimental bond lengths for small to medium size molecules. As the ground states of these molecules at their 
equilibrium geometries can be well described by single-reference wave functions, TAO-DFAs are found to perform very similarly (see Table \ref{table:EXTS}) to the corresponding KS-DFAs \cite{supp}.

\subsection{Dissociation of H$_2$ and N$_2$} 

Due to the presence of strong static correlation effects, the dissociation of H$_2$ and N$_2$ remains an important and challenging subject in KS-DFT. Based on the symmetry constraint, the difference between the 
spin-restricted and spin-unrestricted dissociation limits calculated by an approximate method can be adopted as a quantitative measure of the static correlation error (SCE) of the method \cite{SciYang,SCE,TAO-DFT}. 
Spin-restricted KS-DFAs, global hybrid functionals, LC hybrid functionals, and DH functionals have been shown to perform very poorly for both H$_2$ and N$_2$ dissociation curves, leading to very large SCEs. By contrast, 
as discussed in Ref.\ \cite{TAO-DFT}, spin-restricted TAO-LDA (with a $\theta$ between 30 and 50 mhartree) can properly dissociate H$_2$ and N$_2$ (yielding vanishingly small SCEs) to the respective spin-unrestricted 
dissociation limits, which is closely related to that the distribution of orbital occupation numbers $\{f_{i,\sigma}\}$ (related to the chosen $\theta$) matches reasonably well with that of the NOONs. 

\begin{figure} 
\caption{\label{fig:h2lda} Potential energy curves (in relative energy) for the ground state of H$_2$, calculated by spin-restricted TAO-LDA (with various $\theta$). 
The zeros of energy are set at the respective spin-unrestricted dissociation limits. The $\theta=0$ case corresponds to spin-restricted KS-LDA.} 
\includegraphics[scale=1.0]{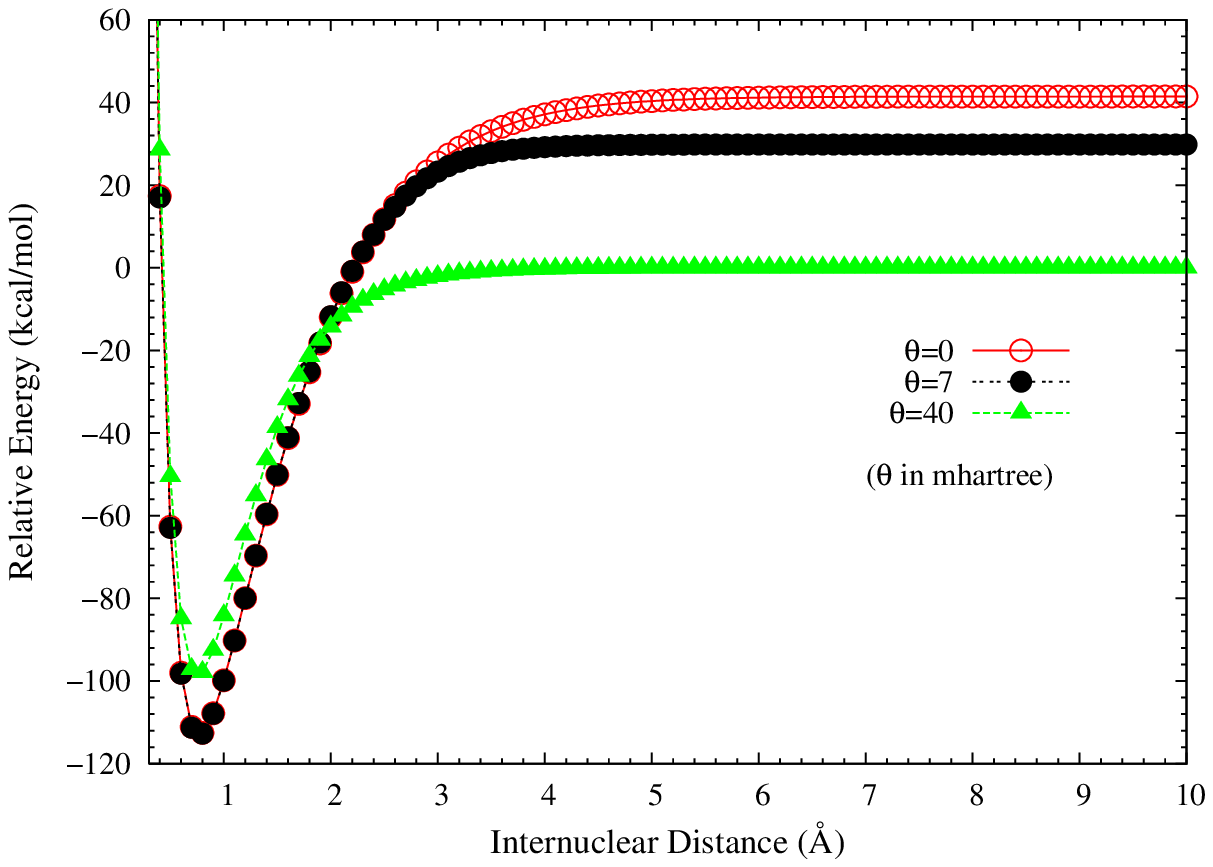} 
\end{figure}

\begin{figure} 
\caption{\label{fig:h2pbe} Same as Fig.\ \ref{fig:h2lda}, but for spin-restricted TAO-PBE (with various $\theta$). The $\theta=0$ case corresponds to spin-restricted KS-PBE.} 
\includegraphics[scale=1.0]{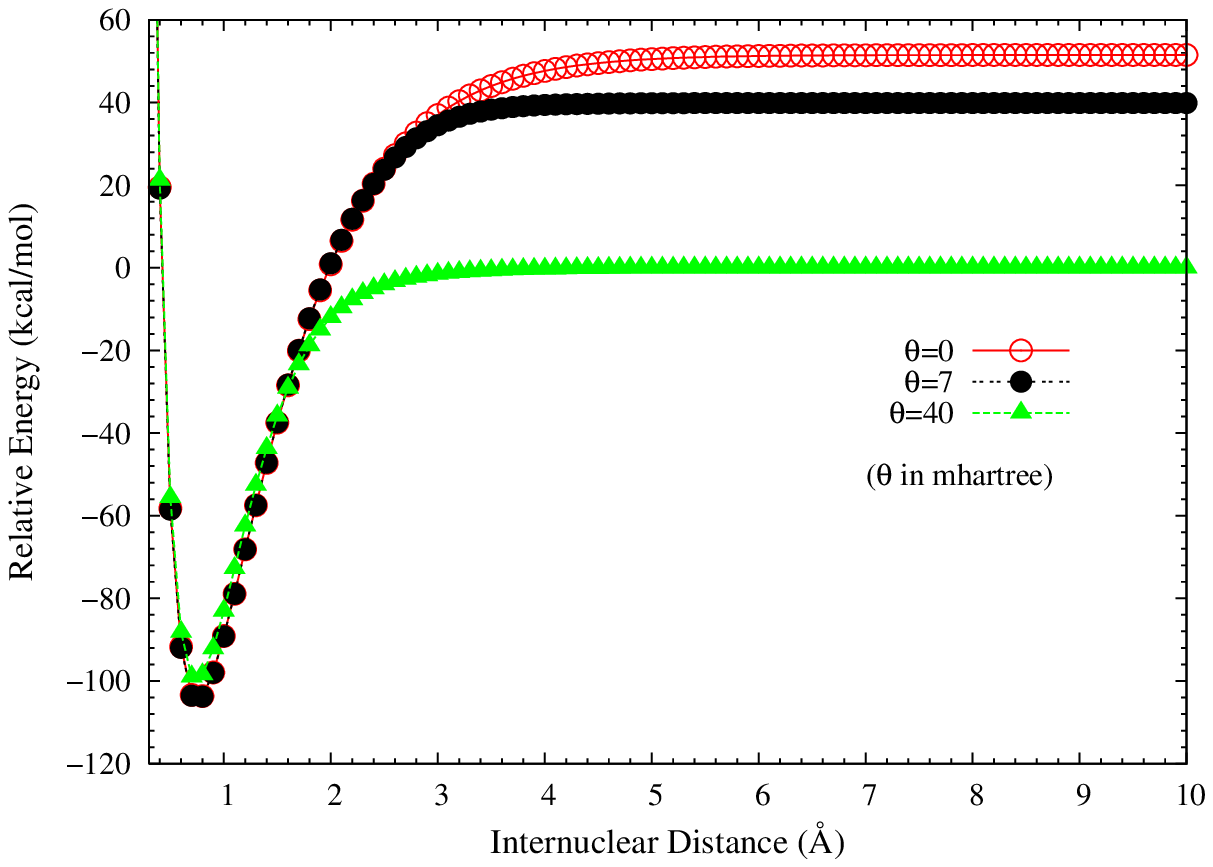} 
\end{figure}

\begin{figure} 
\caption{\label{fig:h2blyp} Same as Fig.\ \ref{fig:h2lda}, but for spin-restricted TAO-BLYP (with various $\theta$). The $\theta=0$ case corresponds to spin-restricted KS-BLYP.} 
\includegraphics[scale=1.0]{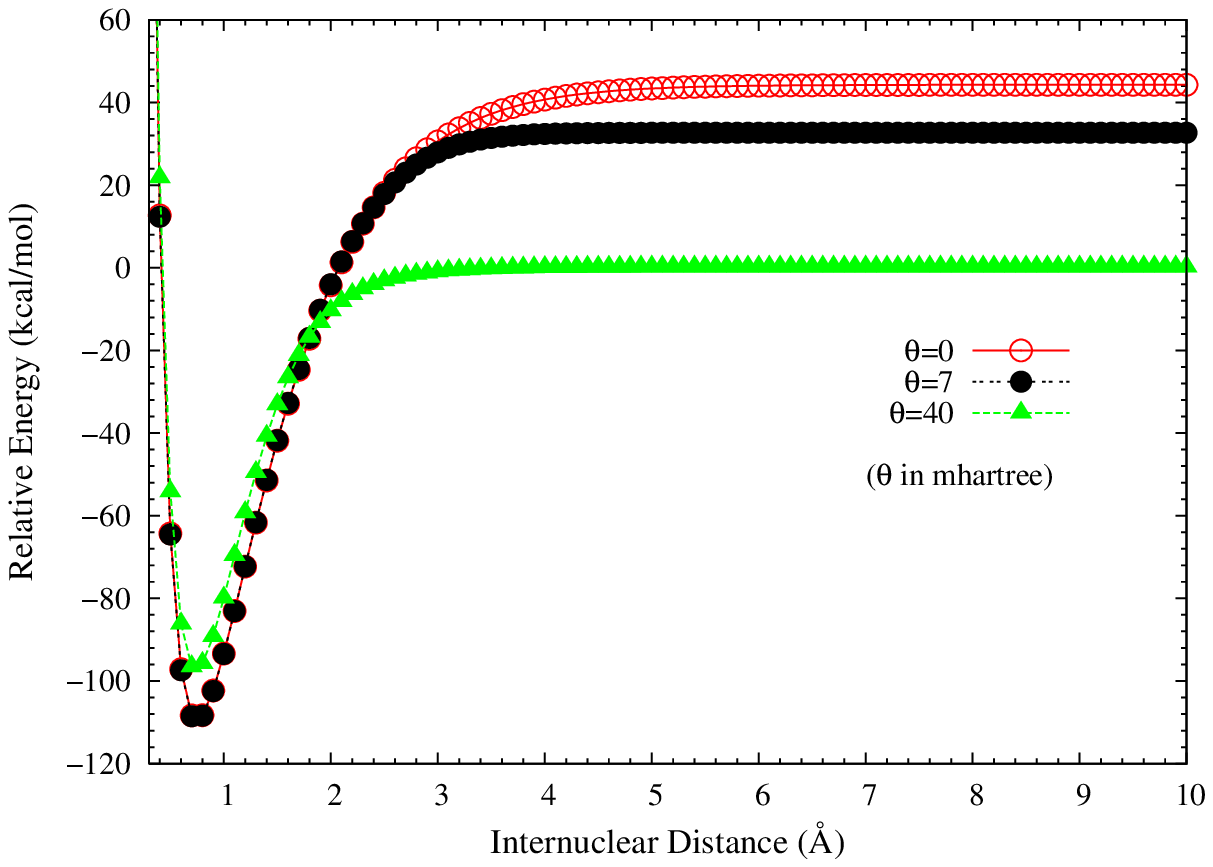} 
\end{figure}

\begin{figure} 
\caption{\label{fig:h2blyp-d} Same as Fig.\ \ref{fig:h2lda}, but for spin-restricted TAO-BLYP-D (with various $\theta$). The $\theta=0$ case corresponds to spin-restricted KS-BLYP-D.} 
\includegraphics[scale=1.0]{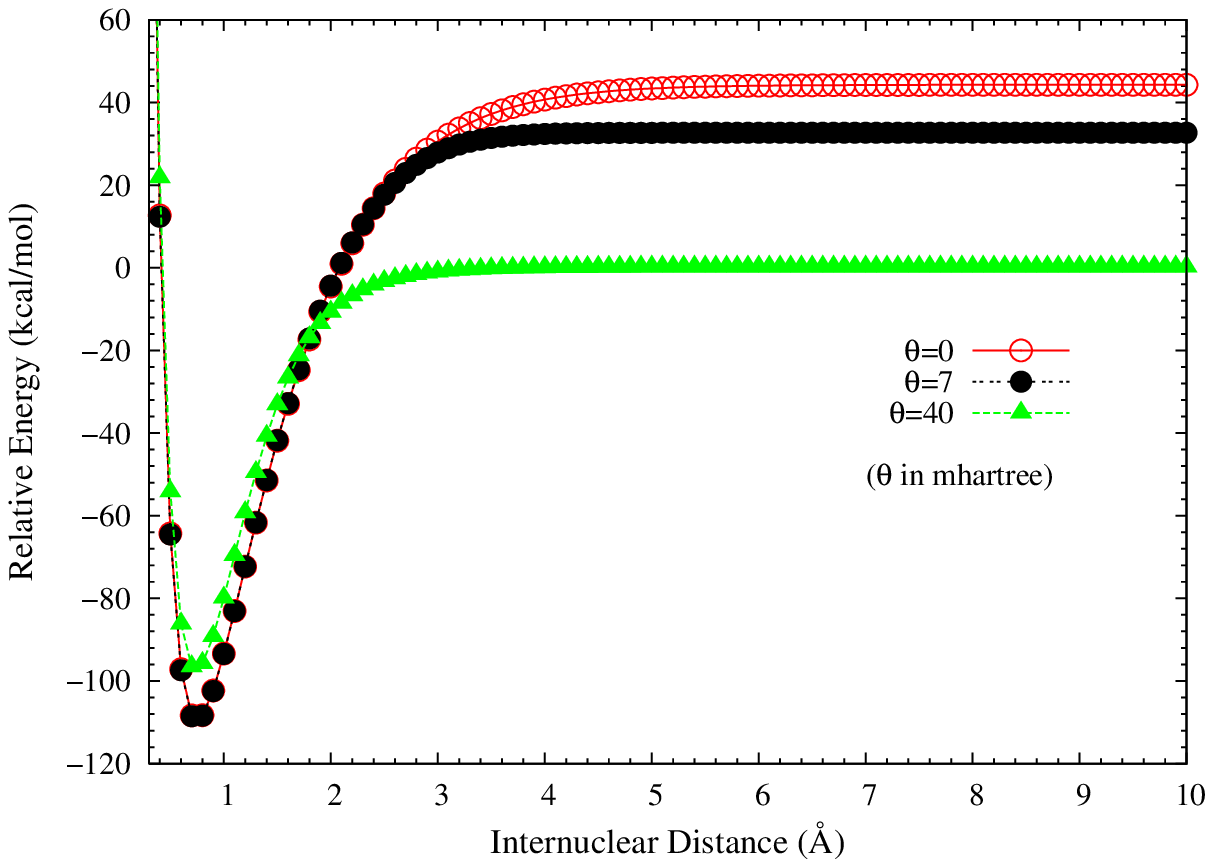} 
\end{figure}

\begin{figure} 
\caption{\label{fig:n2lda} Potential energy curves (in relative energy) for the ground state of N$_2$, calculated by spin-restricted TAO-LDA (with various $\theta$). 
The zeros of energy are set at the respective spin-unrestricted dissociation limits. The $\theta=0$ case corresponds to spin-restricted KS-LDA.} 
\includegraphics[scale=1.0]{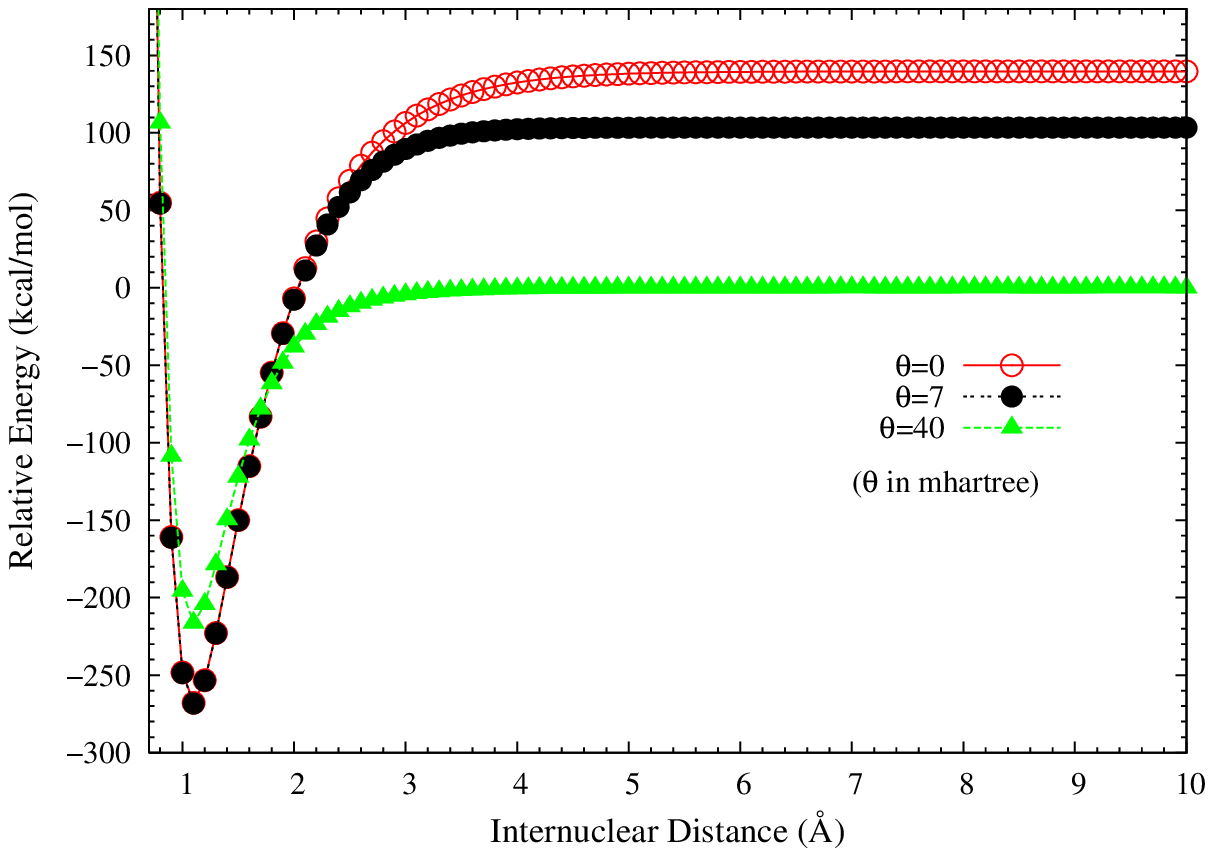} 
\end{figure}

\begin{figure} 
\caption{\label{fig:n2pbe} Same as Fig.\ \ref{fig:n2lda}, but for spin-restricted TAO-PBE (with various $\theta$). The $\theta=0$ case corresponds to spin-restricted KS-PBE.} 
\includegraphics[scale=1.0]{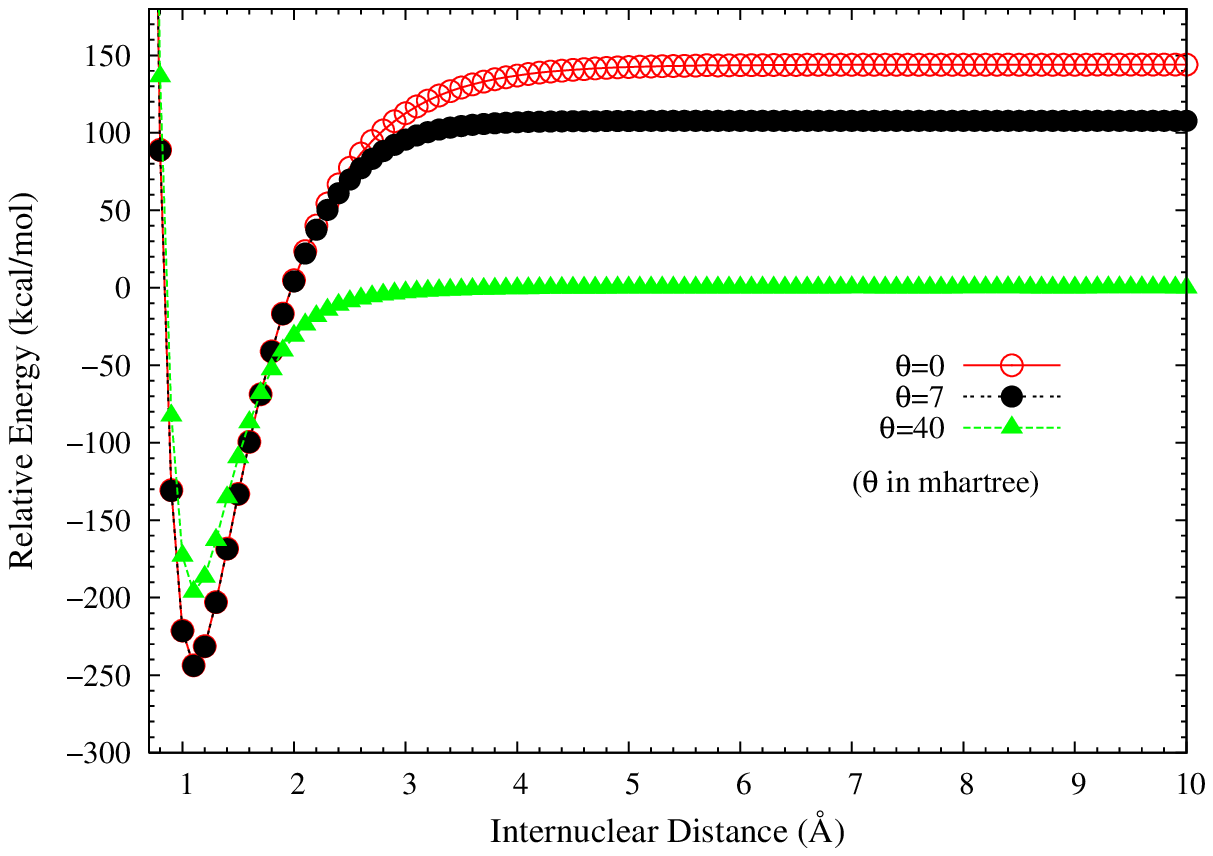} 
\end{figure}

\begin{figure} 
\caption{\label{fig:n2blyp} Same as Fig.\ \ref{fig:n2lda}, but for spin-restricted TAO-BLYP (with various $\theta$). The $\theta=0$ case corresponds to spin-restricted KS-BLYP.} 
\includegraphics[scale=1.0]{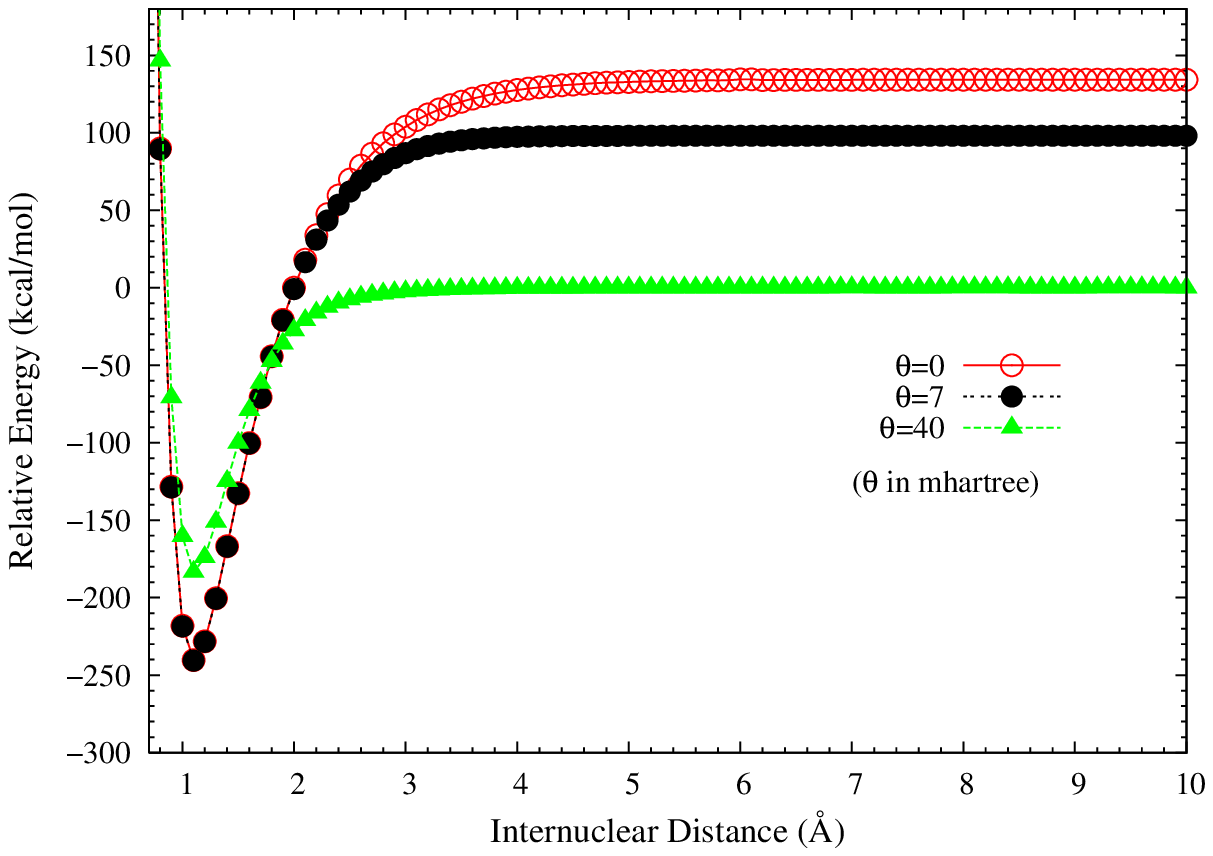} 
\end{figure}

\begin{figure} 
\caption{\label{fig:n2blyp-d} Same as Fig.\ \ref{fig:n2lda}, but for spin-restricted TAO-BLYP-D (with various $\theta$). The $\theta=0$ case corresponds to spin-restricted KS-BLYP-D.} 
\includegraphics[scale=1.0]{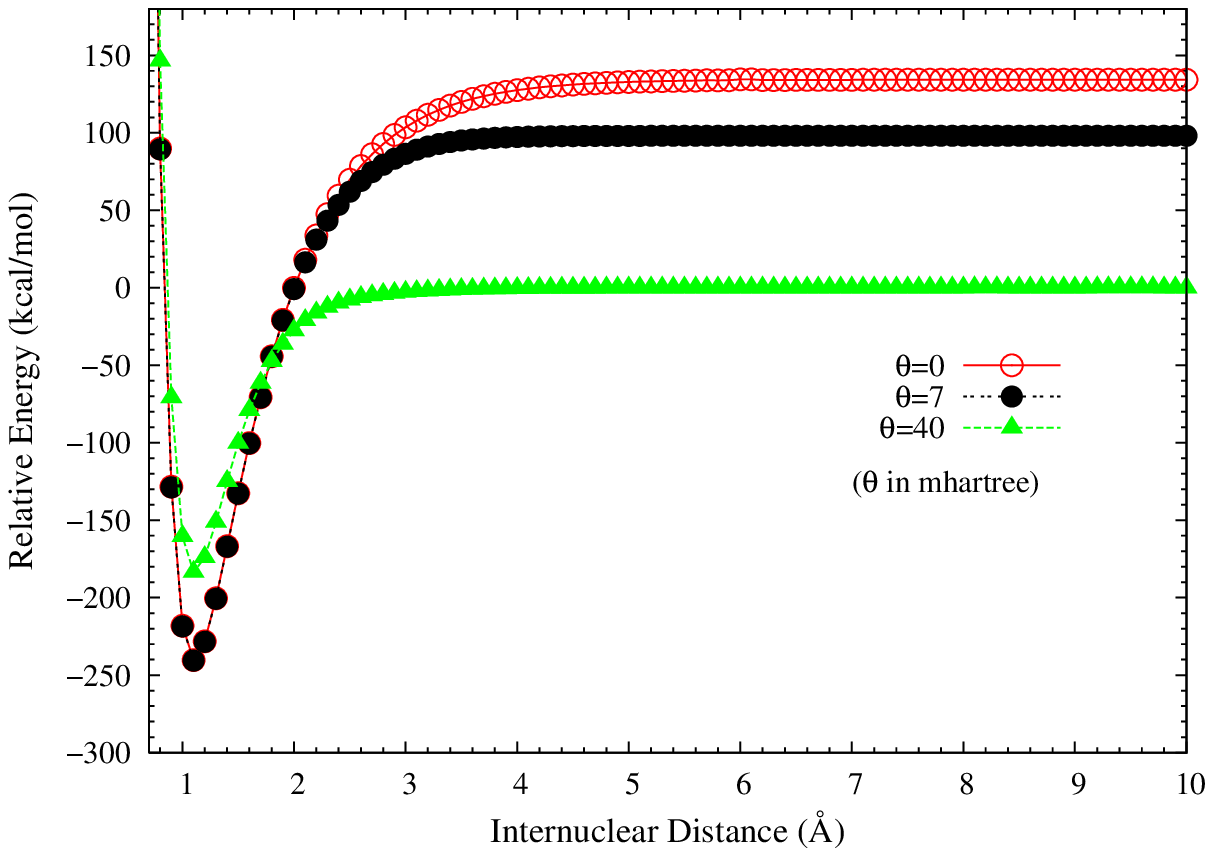} 
\end{figure}

To examine the performance of the present method, the potential energy curves (in relative energy) for the ground state of H$_2$, calculated by spin-restricted TAO-DFAs (with various $\theta$), are shown in 
Figs.\ \ref{fig:h2lda}, \ref{fig:h2pbe}, \ref{fig:h2blyp}, and \ref{fig:h2blyp-d}, where the zeros of energy are set at the respective spin-unrestricted dissociation limits. Near the equilibrium geometry of H$_2$, where the 
single-reference character is dominant, TAO-DFAs (with $\theta$ = 7 mhartree) perform very similarly to the corresponding KS-DFAs (the $\theta=0$ cases). At the dissociation limit, where the multi-reference character 
becomes pronounced, TAO-DFAs (with $\theta$ = 40 mhartree) have vanishingly small SCEs, while TAO-DFAs (with $\theta$ = 7 mhartree) and KS-DFAs have noticeable SCEs. 
Overall, the SCEs of TAO-DFAs are not sensitive to the choice of DFAs (mainly responsible for short-range XC effects), but sensitive to the value of $\theta$ (closely related to the distribution of the NOONs). 
Similar results are also found for N$_2$ dissociation, as shown in Figs.\ \ref{fig:n2lda}, \ref{fig:n2pbe}, \ref{fig:n2blyp}, and \ref{fig:n2blyp-d}.

\section{Electronic Properties of Linear Acenes} 

\begin{figure}
\caption{\label{fig:6-acene} Hexacene, consisting of 6 linearly fused benzene rings, is designated as 6-acene.} 
\includegraphics[scale=0.5]{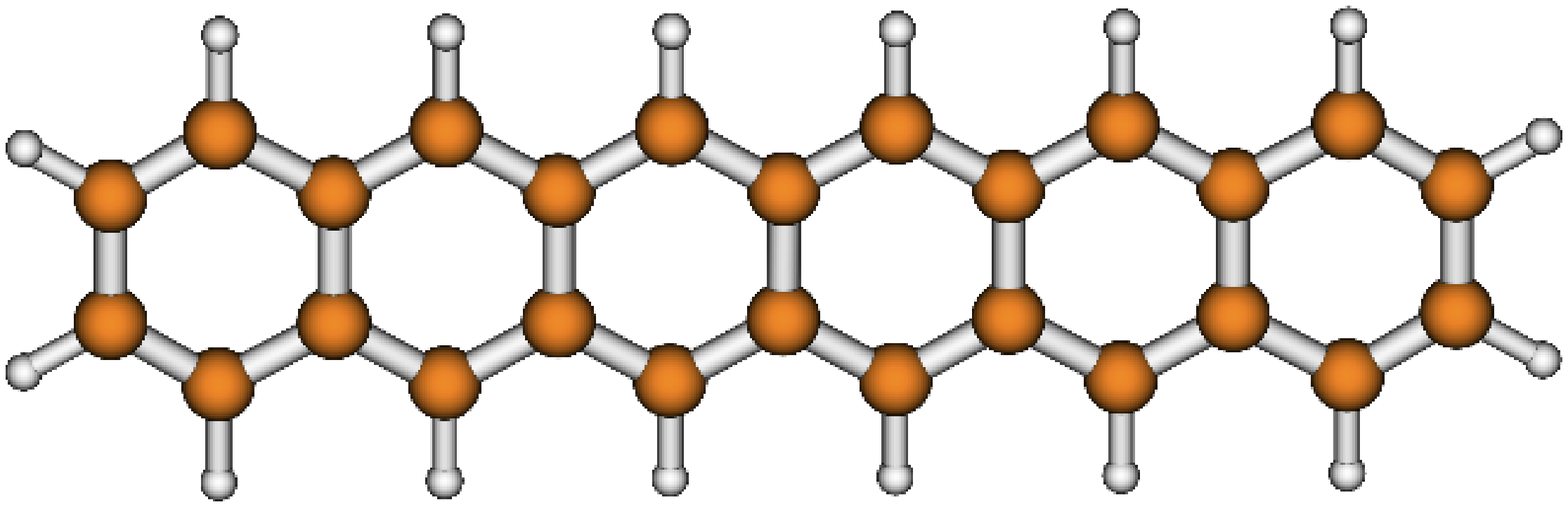}
\end{figure}

Linear $n$-acenes (C$_{4n+2}$H$_{2n+4}$), consisting of $n$ linearly fused benzene rings (see Fig.\ \ref{fig:6-acene}), have recently attracted significant attention from many experimental and theoretical researchers 
due to their fascinating electronic properties, opening up tremendous possibilities to realize electronic and spintronic 
nanodevices \cite{2-acene,3-acene,4-acene,5-acene,aceneTH,aceneIP,aceneChan,acene_IPEAFG,aceneJiang,aceneEA,aceneHajgato,aceneHajgato2,aceneMazziotti,TAO-DFT,entropy}. The electronic properties of acenes 
largely depend on the chain lengths. Despite increasing interest in acenes, it remains very challenging to explore the properties of long-chain acenes from both experimental and theoretical perspectives. 
Experimentally, the difficulties in synthesizing long-chain acenes and their instability following isolation have been attributed to their radical character. 
Accordingly, the experimental singlet-triplet energy gaps (ST gaps) of $n$-acenes are only available up to pentacene \cite{2-acene,3-acene,4-acene,5-acene}. 
On the theoretical side, acenes, which belong to conjugated $\pi$-orbital systems, typically require high-level {\it ab initio} multi-reference methods, such as the DMRG algorithm \cite{aceneChan}, 
the variational two-electron reduced density matrix (2-RDM) method \cite{aceneMazziotti}, or other high-level methods \cite{aceneIP,aceneEA,aceneHajgato,aceneHajgato2}, to capture the essential strong static 
correlation effects. However, these methods are prohibitively expensive for the study of large acenes. 

As mentioned previously, for systems with pronounced strong static correlation effects, such as large acenes, the predictions from KS-DFAs can be problematic \cite{SciYang,RevYang,TAO-DFT}. 
To examine how TAO-DFAs improve upon KS-DFAs here, spin-unrestricted KS-DFT and TAO-DFT calculations, employing various XC functionals: LDA, PBE, BLYP, and BLYP-D, are performed, 
using the 6-31G(d) basis set (up to 100-acene), for the lowest singlet and triplet energies on the respective geometries that were fully optimized at the same level. The ST gap of $n$-acene is calculated as 
$(E_{\text{T}} - E_{\text{S}})$, the energy difference between the lowest triplet (T) and singlet (S) states of $n$-acene. 

\begin{figure}
\caption{\label{fig:stg} Singlet-triplet energy gap as a function of the acene length, calculated by various functionals in spin-unrestricted KS-DFT and TAO-DFT. 
The experimental data (uncorrected for zero-point vibrations, thermal vibrations, etc.) are taken from Refs.\ \cite{2-acene,3-acene,4-acene,5-acene}, the DMRG data are taken from Ref.\ \cite{aceneChan}, 
and the CCSD(T)/CBS data are taken from Ref.\ \cite{aceneHajgato2}.} 
\includegraphics[scale=1.0]{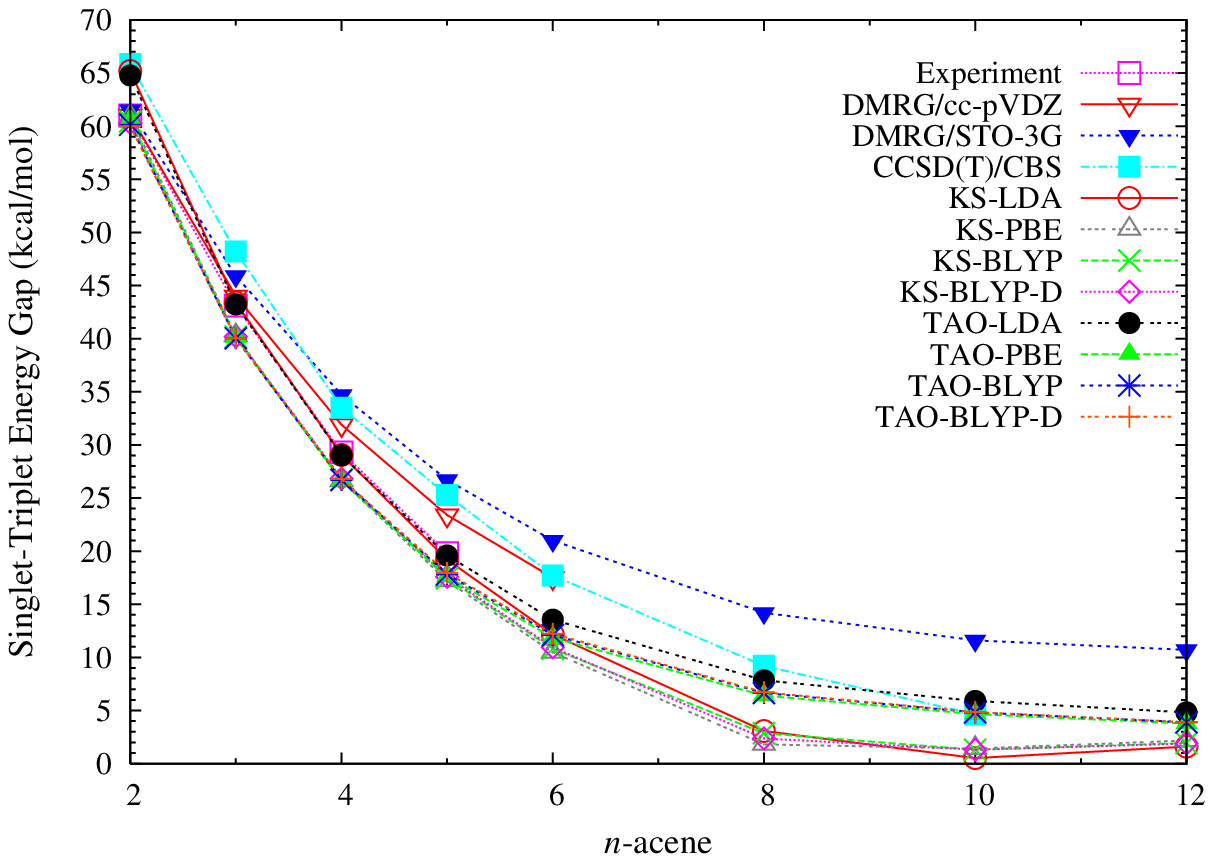}
\end{figure}

\begin{figure}
\caption{\label{fig:stg2} Same as Fig.\ {\ref{fig:stg}}, but for the larger acenes.} 
\includegraphics[scale=1.0]{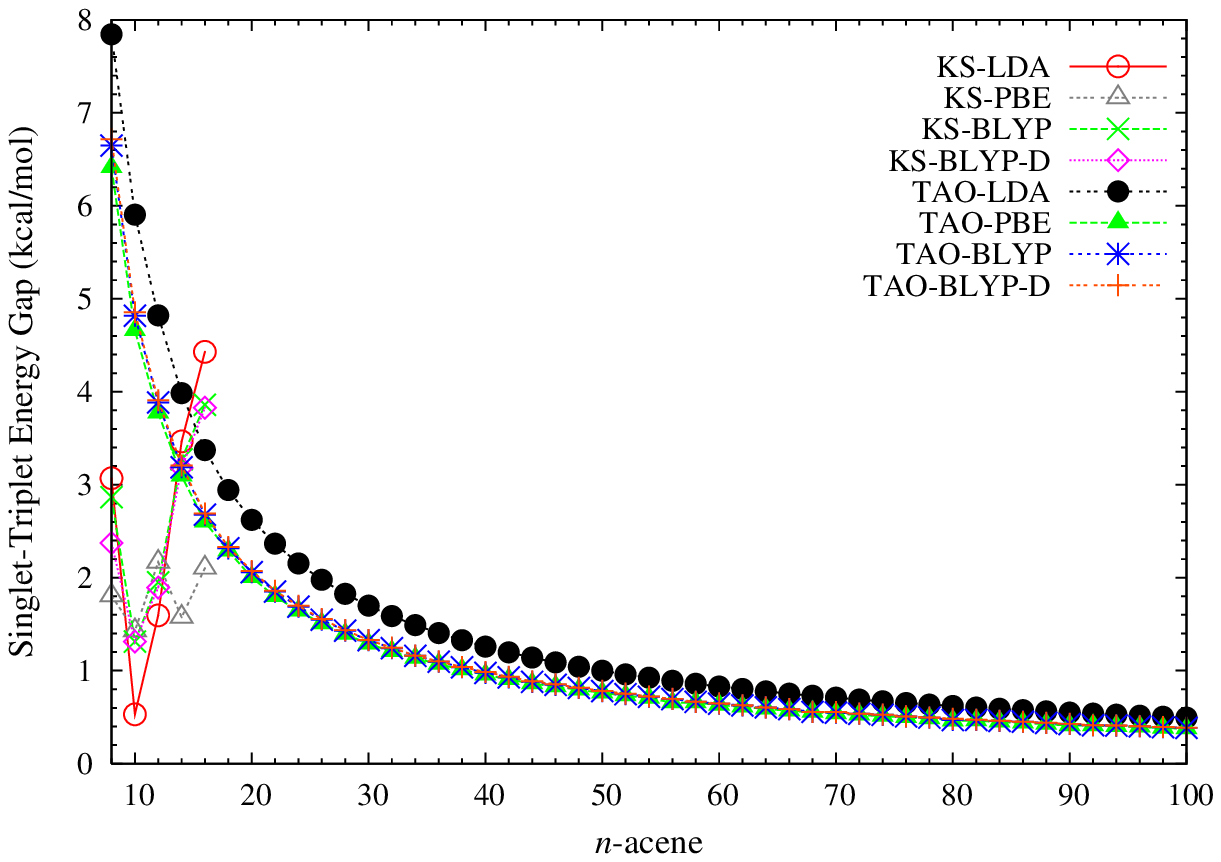}
\end{figure}

As shown in Figs.\ \ref{fig:stg} and \ref{fig:stg2}, in contrast to the DMRG results \cite{aceneChan}, the ST gaps calculated by spin-unrestricted KS-DFT, unexpectedly increase beyond 10-acene, due to unphysical 
symmetry-breaking effects \cite{supp}. By contrast, the ST gaps calculated by spin-unrestricted TAO-DFT, which are in good agreement with the existing experimental and high-level {\it ab initio} 
data \cite{aceneChan,aceneHajgato2}, are shown to decrease monotonically with the increase of chain length. This shows that the ground states of acenes are singlets for all the chain lengths studied. 
The ST gap of the largest acene studied here (100-acene) is 0.38 kcal/mol for TAO-PBE and TAO-BLYP, 0.39 kcal/mol for TAO-BLYP-D, and 0.49 kcal/mol for TAO-LDA \cite{supp}. 
To examine the possible symmetry-breaking effects, spin-restricted TAO-DFT calculations are also performed for the lowest singlet energies on the respective geometries that were fully optimized at the same level. 
The spin-unrestricted and spin-restricted TAO-DFT calculations are found to essentially yield the same energy value for the lowest singlet state of $n$-acene (i.e., no unphysical symmetry-breaking effects). 

\begin{figure}
\caption{\label{fig:ip} Vertical ionization potential for the lowest singlet state of $n$-acene as a function of the acene length, calculated by various functionals in spin-unrestricted TAO-DFT. 
The experimental data are taken from the compilation in Ref.\ \cite{acene_IPEAFG}, and the CCSD(T)/CBS data are taken from Ref.\ \cite{aceneIP}.} 
\includegraphics[scale=1.0]{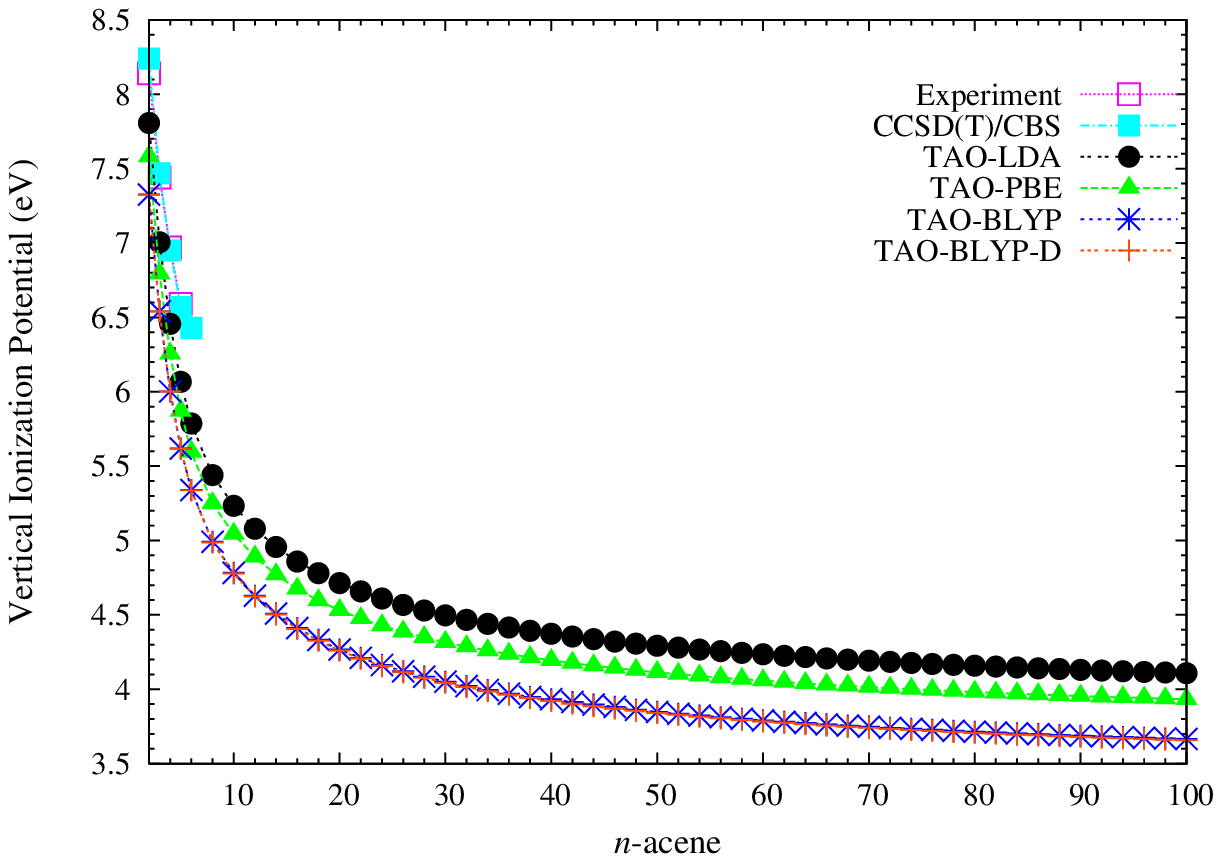} 
\end{figure}

\begin{figure}
\caption{\label{fig:ea} Vertical electron affinity for the lowest singlet state of $n$-acene as a function of the acene length, calculated by various functionals in spin-unrestricted TAO-DFT. 
The experimental data are taken from the compilation in Ref.\ \cite{acene_IPEAFG}, and the CCSD(T)/CBS data are taken from Ref.\ \cite{aceneEA}.} 
\includegraphics[scale=1.0]{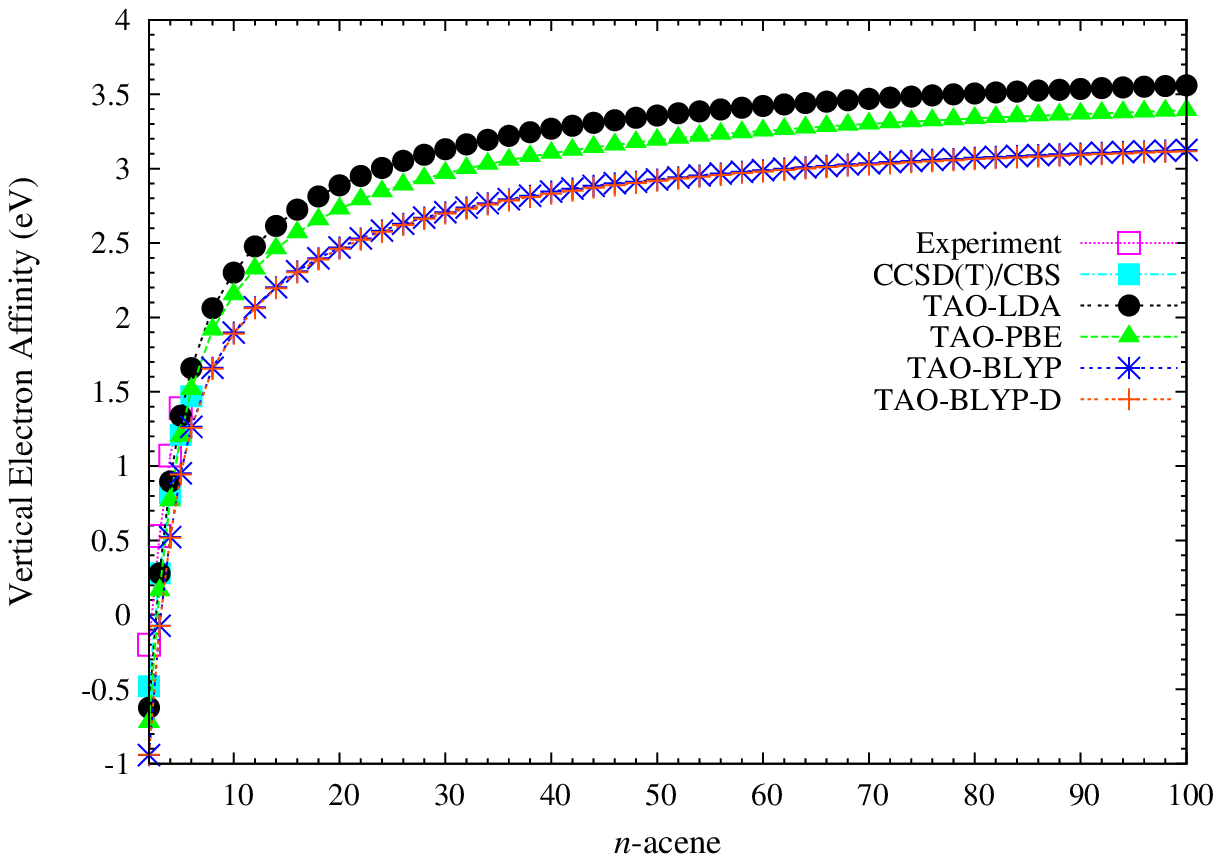}
\end{figure}

\begin{figure}
\caption{\label{fig:fg} Fundamental gap for the lowest singlet state of $n$-acene as a function of the acene length, calculated by various functionals in spin-unrestricted TAO-DFT. 
The experimental data are taken from the compilation in Ref.\ \cite{acene_IPEAFG}, and the CCSD(T)/CBS data are taken from Refs.\ \cite{aceneIP,aceneEA}.} 
\includegraphics[scale=1.0]{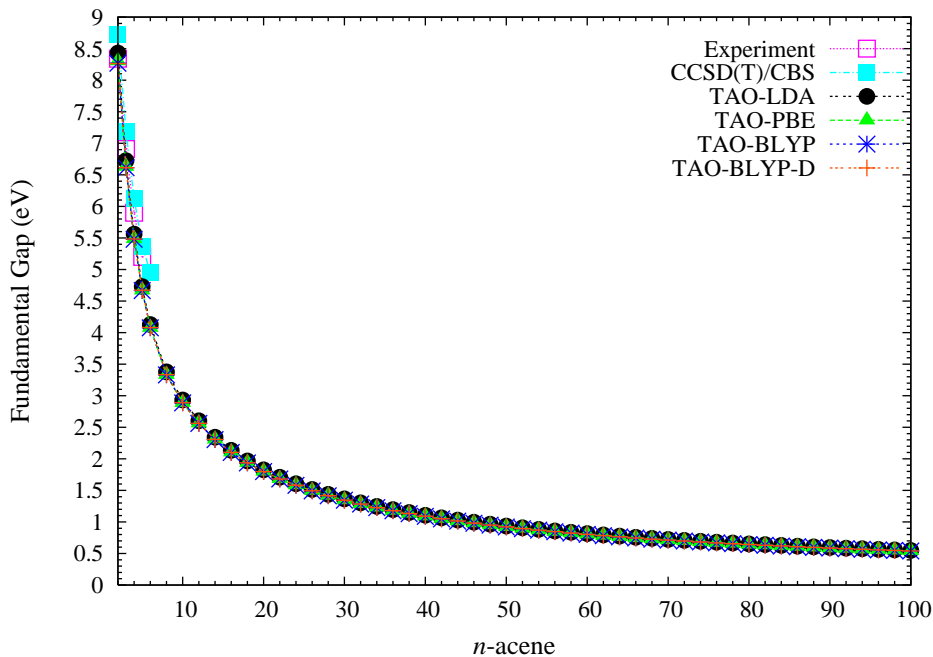} 
\end{figure}

At the optimized geometry of the lowest singlet state (i.e., the ground state) of $n$-acene, containing $N$ electrons, the vertical ionization potential $\text{IP}_{v}$ and the vertical electron affinity $\text{EA}_{v}$ can be 
calculated by 
\begin{equation}\label{eq:ip} 
\text{IP}_{v} = {E}_{N-1} - {E}_{N}, 
\end{equation} 
and 
\begin{equation}\label{eq:ea} 
\text{EA}_{v} = {E}_{N} - {E}_{N+1}, 
\end{equation} 
respectively, with ${E}_{N}$ being the total energy of the $N$-electron system. Accordingly, the fundamental gap $E_{g}$ can be calculated by 
\begin{equation}\label{eq:fg} 
E_{g} = \text{IP}_{v} - \text{EA}_{v} = {E}_{N+1} + {E}_{N-1} - 2{E}_{N}. 
\end{equation} 
As the size of the acene increases, $\text{IP}_{v}$ (see Fig.\ \ref{fig:ip}) monotonically decreases and $\text{EA}_{v}$ (see Fig.\ \ref{fig:ea}) monotonically increases, yielding a monotonically deceasing 
$E_{g}$ (see Fig.\ \ref{fig:fg}). In contrast to the calculated $\text{IP}_{v}$ and $\text{EA}_{v}$, $E_{g}$ seems to be rather insensitive to the choice of the XC functionals in TAO-DFT. 
For 100-acene, $E_{g}$ is 0.54 eV for TAO-PBE, TAO-BLYP, and TAO-BLYP-D, and 0.55 eV for TAO-LDA. 
Note that the calculated $E_{g}$ is within the most interesting range (1 to 3 eV) for $n$-acene ($n$: 44 to 10), giving promise for applications of acenes in nanoelectronics \cite{supp}. 

\begin{figure}
\caption{\label{fig:entropy} Symmetrized von Neumann entropy for the lowest singlet state of $n$-acene as a function of the acene length, calculated by various functionals in spin-restricted TAO-DFT.} 
\includegraphics[scale=1.0]{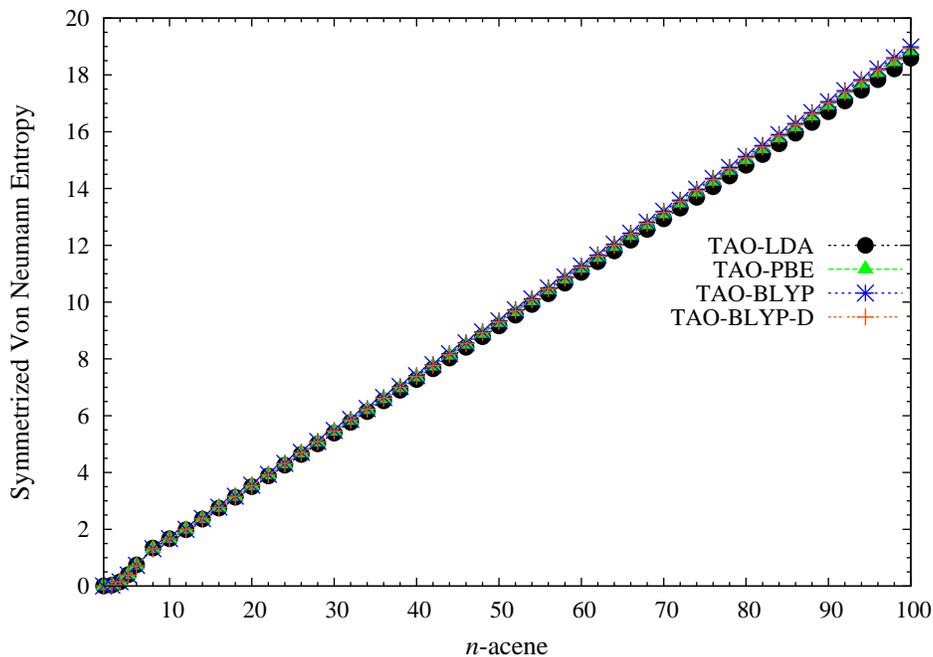}
\end{figure}

The orbital occupation numbers in TAO-DFT provide information useful in assessing the possible polyradical character of $n$-acene. As they are closely related to the NOONs \cite{NO}, we calculate 
the symmetrized von Neumann entropy (e.g., see Eq.\ (9) of Ref.\ \cite{entropy}) 
\begin{equation}\label{eq:svn} 
S_{\text{vN}} = -\frac{1}{2} \sum_{\sigma=\alpha,\beta} \sum_{i=1}^{\infty} \bigg\lbrace f_{i,\sigma}\ \text{ln}(f_{i,\sigma}) + (1-f_{i,\sigma})\ \text{ln}(1-f_{i,\sigma}) \bigg\rbrace, 
\end{equation} 
for the lowest singlet state of $n$-acene as a function of the acene length, by spin-restricted TAO-DFT \cite{supp}. Here $S_{\text{vN}}$ essentially provides no contributions for a single-reference system ($\{f_{i,\sigma}\}$ 
are close to either 0 or 1), and quickly increases with the number of active orbitals ($\{f_{i,\sigma}\}$ are fractional for active orbitals, and are close to either 0 or 1 for others). Note that $S_{\text{vN}}$, which is simply 
$(-\frac{1}{2 \theta} E_{S}^{\theta}[\{f_{i,\alpha} \}, \{f_{i,\beta} \}])$, can be readily obtained in TAO-DFT. As shown in Fig.\ \ref{fig:entropy}, $S_{\text{vN}}$, which is closely related to the polyradical character of $n$-acene, 
increases monotonically with the chain length, supporting the previous finding that large acenes should exhibit polyradical character \cite{aceneChan,aceneJiang}.

\section{Conclusions} 

We have proposed generalized-gradient approximations to TAO-DFT. The resulting TAO-GGAs have been shown to significantly outperform our previous TAO-LDA for a wide range of applications, 
such as thermochemistry, kinetics, and reaction energies. For noncovalent interactions, TAO-GGAs with empirical dispersion corrections have been shown to yield excellent performance. 
Due to their computational efficiency, TAO-LDA and TAO-GGAs have been applied to study the electronic properties of acenes, including the ST gaps, vertical ionization potentials, vertical electron affinities, 
fundamental gaps, and symmetrized von Neumann entropy (i.e., a measure of polyradical character). The ground states of acenes have been shown to be singlets for all the chain lengths studied here. 
With the increase of acene length, the ST gaps, vertical ionization potentials, and fundamental gaps decrease monotonically, while the vertical electron affinities and polyradical character increase monotonically. 

Although only three GGAs (PBE, BLYP, and BLYP-D) are examined in this work, the good properties of various GGAs (e.g., those recently developed for broad applicability or for 
specific properties \cite{GGA1,GGA2,GGA3,GGA4,GGA5,GGA6,GGA7}) in KS-DFT are expected to be preserved in TAO-DFT (with a sufficiently small $\theta$ = 7 mhartree). However, for some 
multi-reference systems (e.g., H$_2$ and N$_2$ dissociation curves), TAO-DFAs (with $\theta$ = 7 mhartree) may not provide a sufficient amount of static correlation energy. 
As a system-dependent $\theta$ (related to the distributions of NOONs) is expected to enhance the performance of TAO-DFAs for a wide range of single- and multi-reference systems, we plan to pursue this in the future.

\begin{acknowledgements} 

This work was supported by the National Science Council of Taiwan (Grant No.\ NSC101-2112-M-002-017-MY3), National Taiwan University (Grant No.\ NTU-CDP-102R7855), 
the Center for Quantum Science and Engineering at NTU (Subprojects:\ 102R891401 and 102R891403), and the National Center for Theoretical Sciences of Taiwan. 
We are grateful to the Computer and Information Networking Center at NTU for the support of high-performance computing facilities. 

\end{acknowledgements}

\bibliographystyle{jcp}

\newpage 

\newpage

\newpage 

\newpage 

\newpage 

\newpage 

\newpage 

\newpage 

\newpage

\newpage

\newpage

\newpage

\newpage

\newpage

\newpage

\newpage
\begin{table*}
\caption{\label{table:training} Statistical errors (in kcal/mol) of the $\omega$B97 training set \cite{wB97X}.} 
\begin{ruledtabular}
\begin{tabular*}{\textwidth}{llrrrrrrrr}
& \multicolumn{4}{r}{KS-DFT} & \multicolumn{4}{r}{TAO-DFT} \\
\cline{3-6}
\cline{7-10}
System & Error & LDA & PBE & BLYP & BLYP-D & LDA & PBE & BLYP & BLYP-D \\
		\hline
		G3/99 & MSE & 120.60 	&	20.90 	&	-4.59 	&	-0.83 	&	95.02 	&	7.91 	&	-16.24 	&	-12.27 	\\ 
		(223) & MAE & 120.60 	&	21.51 	&	9.76 	&	7.03 	&	95.04 	&	11.41 	&	19.01 	&	15.33 	\\
		           & rms & 142.51 	&	26.30 	&	12.96 	&	9.17 	&	114.19 	&	15.07 	&	24.24 	&	19.35 	\\
		\hline
		IP     & MSE & 3.42 	&	0.03 	&	-1.50 	&	-1.50 	&	1.79 	&	-1.08 	&	-2.61 	&	-2.61 	\\
		(40)  & MAE & 5.54 	&	3.46 	&	4.43 	&	4.44 	&	6.18 	&	4.86 	&	6.10 	&	6.10 	\\
		          & rms & 6.66 	&	4.35 	&	5.28 	&	5.29 	&	7.63 	&	6.00 	&	7.40 	&	7.40 	\\		          
		\hline
		EA	& MSE & 6.45 	&	1.72 	&	0.36 	&	0.36 	&	4.20 	&	0.22 	&	-1.08 	&	-1.07 	\\
		(25)  & MAE & 6.45 	&	2.42 	&	2.57 	&	2.57 	&	5.49 	&	2.88 	&	4.38 	&	4.40 	\\
		         & rms & 7.29 	&	3.06 	&	3.17 	&	3.17 	&	6.45 	&	3.44 	&	5.44 	&	5.47 	\\
		\hline
		PA  & MSE & -5.91 	&	-0.83 	&	-1.47 	&	-1.09 	&	-5.66 	&	-0.58 	&	-1.22 	&	-0.84 	\\
		(8)  & MAE & 5.91 	&	1.60 	&	1.58 	&	1.55 	&	5.66 	&	1.47 	&	1.50 	&	1.55 	\\
		       & rms & 6.40 	&	1.91 	&	2.10 	&	1.98 	&	6.16 	&	1.80 	&	1.94 	&	1.86 	\\
		\hline
		NHTBH & MSE & -12.41 	&	-8.52 	&	-8.69 	&	-9.32 	&	-11.93 	&	-8.38 	&	-8.52 	&	-9.15 	\\
		(38)        & MAE & 12.62 	&	8.62 	&	8.72 	&	9.35 	&	12.15 	&	8.49 	&	8.56 	&	9.19 	\\
		               & rms & 16.13 	&	10.61 	&	10.27 	&	10.83 	&	15.09 	&	10.28 	&	9.90 	&	10.46 	\\
		\hline
		HTBH & MSE & -17.90 	&	-9.67 	&	-7.84 	&	-8.89 	&	-16.34 	&	-9.20 	&	-7.25 	&	-8.33 	\\
		(38)     & MAE & 17.90 	&	9.67 	&	7.84 	&	8.89 	&	16.34 	&	9.20 	&	7.29 	&	8.34 	\\
		            & rms & 18.92 	&	10.37 	&	8.66 	&	9.52 	&	17.06 	&	9.87 	&	8.24 	&	9.17 	\\
		\hline
		S22	& MSE & -1.97 	&	2.77 	&	5.05 	&	0.23 	&	-2.30 	&	2.44 	&	4.70 	&	-0.12 	\\
		(22)  & MAE & 2.08 	&	2.77 	&	5.05 	&	0.33 	&	2.33 	&	2.44 	&	4.70 	&	0.28 	\\
		         & rms & 3.18 	&	3.89 	&	6.31 	&	0.45 	&	3.40 	&	3.57 	&	5.95 	&	0.37 	\\
		\hline
		Total	 & MSE & 65.86 	&	10.33 	&	-4.07 	&	-2.37 	&	51.26 	&	2.81 	&	-10.81 	&	-8.99 	\\
		(394) & MAE & 72.41 	&	14.63 	&	8.05 	&	6.40 	&	57.76 	&	9.01 	&	13.48 	&	11.31 	\\
		           & rms & 107.53 	&	20.40 	&	10.88 	&	8.44 	&	86.26 	&	12.38 	&	18.92 	&	15.43 	\\
\end{tabular*}
\end{ruledtabular}
\end{table*}

\newpage
\begin{table*}
\begin{center}
\caption{\label{table:reall} Statistical errors (in kcal/mol) of the reaction energies of the 30 chemical reactions in the NHTBH38/04 and HTBH38/04 sets \cite{BH}.} 
\begin{ruledtabular}
\begin{tabular}{lrrrrrrrr}
& \multicolumn{3}{r}{KS-DFT} & \multicolumn{4}{r}{TAO-DFT} \\
\cline{2-5}
\cline{6-9}
& LDA & PBE & BLYP & BLYP-D & LDA & PBE & BLYP & BLYP-D \\
\hline
MSE&       	-0.41 	&	1.08 	        &	0.80 	       &	0.74 	       &	-1.32 	&	0.23 	       &	-0.12 	&	-0.20 	\\
MAE&	         8.51 	&	4.39 	        &	3.23 	       &	3.02 	       &	7.09 	       &	3.97 	       &	3.80 	       &	3.67 	\\
rms&                11.10 	&	6.24 	        &	4.37 	       &	4.20 	       &	9.38 	       &	5.97 	       &	4.95 	       &	4.89 	\\
Max($-$)& 	-18.31 	&	-7.86 	&	-7.24 	&	-7.28 	&	-15.92 	&	-8.89 	&	-11.24 	&	-11.71 	\\
Max($+$)&	35.68 	&	22.59 	&	11.96 	&	12.03 	&	30.50 	&	21.60 	&	10.65 	&	10.73 	\\
\end{tabular}
\end{ruledtabular}
\end{center}
\end{table*}

\newpage
\begin{table*}
\begin{center}
\caption{\label{table:EXTS} Statistical errors (in {\AA}) of EXTS \cite{EXTS}.} 
\begin{ruledtabular}
\begin{tabular}{lrrrrrrrr}
& \multicolumn{3}{r}{KS-DFT} & \multicolumn{4}{r}{TAO-DFT} \\
\cline{2-5}
\cline{6-9}
& LDA & PBE & BLYP & BLYP-D & LDA & PBE & BLYP & BLYP-D \\ 
\hline
MSE&	0.004 	&	0.014 	&	0.018 	&	0.019 	&	0.005 	&	0.014 	&	0.019 	&	0.020 	\\
MAE&	0.013 	&	0.015 	&	0.019 	&	0.020 	&	0.013 	&	0.015 	&	0.020 	&	0.021 	\\
rms&       0.017 	&	0.019 	&	0.024 	&	0.025 	&	0.017 	&	0.020 	&	0.025 	&	0.026 	\\
Max($-$)& 	-0.091 	&	-0.069 	&	-0.064 	&	-0.063 	&	-0.091 	&	-0.069 	&	-0.064 	&	-0.064 	\\
Max($+$)&	0.081 	&	0.095 	&	0.103 	&	0.103 	&	0.080 	&	0.095 	&	0.102 	&	0.102 	\\
\end{tabular}
\end{ruledtabular}
\end{center}
\end{table*}

\end{document}